\documentclass[12pt, a4paper]{article}
\usepackage{mlmodern, microtype} % also try mathpazo!
\usepackage{amssymb, amsmath, amsthm}
\usepackage{xcolor, mathtools, mathpartir}
\usepackage[runin]{abstract}
\usepackage[hidelinks]{hyperref}
\usepackage[numbers]{natbib}
\usepackage[nameinlink]{cleveref}

\theoremstyle{definition}
\newtheorem{remark}{Remark}

\newcommand*{\doi}[1]{doi: \href{https://doi.org/#1}{\nolinkurl{#1}}}
\makeatletter
\def\NAT@spacechar{~}
\newcommand*{\bible}[1]{\hyper@link{cite}{cite.sterling:2021:thesis}{\textbf{(#1)}}}
\makeatother
\newcommand{\astr}{\textasteriskcentered}

\title{Normal forms in cubical type theory}
\author{Xu Huang}
\date{}
\begin{document}
\maketitle

\setlength{\abstitleskip}{-\parindent}
\renewcommand{\abstractname}{Abstract.}
\begin{abstract}
This note documents the specification of normal forms in cubical type theory.
The definition is already present in the proof of normalization
for cubical type theory~\cite{sterling:2021:cubical,sterling:2021:thesis},
but we present it in a more traditional style explicitly for reference.
\end{abstract}

\newcommand{\bind}{\mathpunct{.}}
\newcommand{\of}{\mathbin{:}}

\section{Introduction}

Cubical type theory is an extension of Martin-L\"of type theory
that supports Voevodsky's \emph{univalence axiom},
and the notion of \emph{higher inductive types},
while maintaining nice syntactic properties such as normalization.

To achieve this, cubical type theory introduces a lot of judgmental equalities
that makes it non-trivial to extend traditional normalization proofs.
The language of \emph{synthetic Tait computability}~\cite{sterling:2021:logrel}
is introduced with the purpose of organizing the cubical normalization proof,
given by \citet{sterling:2021:cubical}, and elaborated by \citet{sterling:2021:thesis}.

While this language makes it more practical to construct and verify the proof,
it may still be of value to write down at least the definition of normal forms
in cubical type theory using a more traditional syntax.
It may also serve as a bridge for readers to understand synthetic Tait computability.

We will be considering Cartesian (a.k.a.\ ABCFHL) cubical type theory~\cite{abcfhl:2021:cubical},
following the presentation by \citet{sterling:2021:thesis},
although it can be adapted to de Morgan (a.k.a.\ CCHM) type theory~\cite{cchm:2018:cubical}.
All bold section numbers such as \textbf{(0.3\astr3)} refer to \citet{sterling:2021:thesis}.

\section{Martin-L\"of type theory}

\newcommand{\Ctx}{\mathrm{Ctx}}
\newcommand{\Tp}{\mathrm{Tp}}
\newcommand{\Tm}{\mathrm{Tm}}

\newcommand{\istype}{\;\mathsf{type}}
\newcommand{\ispretype}{\;\mathsf{pretype}}
\newcommand{\isvar}{\;\mathsf{\color{gray}var}}
\newcommand{\isne}{\;\mathsf{\color[rgb]{0.8,0.3,0.1}ne}}
\newcommand{\Tmne}{\mathrm{Tm}_{\mathsf{\color[rgb]{0.8,0.3,0.1}ne}}}
\newcommand{\Tpne}{\mathrm{Tp}_{\mathsf{\color[rgb]{0.8,0.3,0.1}ne}}}
\newcommand{\isnf}{\;\mathsf{\color[rgb]{0.2,0.4,0.9}nf}}
\newcommand{\Tmnf}{\mathrm{Tm}_{\mathsf{\color[rgb]{0.2,0.4,0.9}nf}}}
\newcommand{\Tpnf}{\mathrm{Tp}_{\mathsf{\color[rgb]{0.2,0.4,0.9}nf}}}

We begin by discussing the normal forms in Martin-L\"of type theory.
Recall the syntax of Martin-L\"of type theory defines
a set of contexts \(\Ctx\),
a set of types \(\Tp(\Gamma)\) for each context \(\Gamma \in \Ctx\),
and a set of terms \(\Tm(\Gamma, A)\) for each context \(\Gamma \in \Ctx\) and type \(A \in \Tp(\Gamma)\),
considered up to judgmental equality.
We conventionally write elements of these sets using turnstiles, a.k.a.\ sequents:
\begin{center}
\begin{tabular}{ccc}
\(\Gamma \in \Ctx\) & \(\iff\) & \(\Gamma \vdash\) \\
\(A \in \Tp(\Gamma)\) & \(\iff\) & \(\Gamma \vdash A \istype\) \\
\(t \in \Tm(\Gamma, A)\) & \(\iff\) & \(\Gamma \vdash t : A\) \\
\end{tabular}
\end{center}
The definition usually proceeds by a collection of inference rules,
which can be seen as a kind of quotient inductive definition.

\begin{remark}
\citet{gratzer:2021:framework} proposed a framework to define type theories
where the notion of contexts is demoted as secondary.
It greatly simplifies the presentation of a large class of type theories,
and is in fact used in the normalization proof.
See \bible{Ch.\,1} for an exposition.
\end{remark}

In older literature, normal forms are sometimes defined as expressions
with no applicable reduction rules.
However, this definition is often unsuitable in type theories with \(\eta\) laws,
and certainly unsuitable for cubical type theory, where a lot of equations are
hard to orient as reduction rules.
(See \bible{\S6.5.1} for a discussion on its problems.)
Instead, normal forms are defined inductively for each type,
and mutually recursively with another set called \emph{neutral forms}.

Our goal is to define sets \(\Tmnf(\Gamma, A)\) of normal terms
and \(\Tpnf(\Gamma)\) of normal types,
where \(\Gamma \in \Ctx\) and \(A \in \Tp(\Gamma)\).
Note that the type for a normal term is a general type.
We will simultaneously define the set of neutral terms \(\Tmne(\Gamma, A)\)
and types \(\Tpne(\Gamma)\).
We similarly write
\begin{center}
\begin{tabular}{ccc}
\(A \in \Tpne(\Gamma)\) & \(\iff\) & \(\Gamma \vdash A \istype \isne\) \\
\(A \in \Tpnf(\Gamma)\) & \(\iff\) & \(\Gamma \vdash A \istype \isnf\) \\
\(t \in \Tmne(\Gamma, A)\) & \(\iff\) & \(\Gamma \vdash t : A \isne\) \\
\(t \in \Tmnf(\Gamma, A)\) & \(\iff\) & \(\Gamma \vdash t : A \isnf\) \\
\end{tabular}
\end{center}
These are also given by a quotient inductive definition,
but the difference is that the equations over normal forms and neutral forms
are much simpler than that of terms and types,
so that determining whether two normal forms are equal is straightforward.
In many cases, there are no equations imposed at all
(apart from $\alpha$ equivalence, if the presentation uses named variables).

\newcommand{\incl}{\operatorname{\uparrow}}
\newcommand{\coll}{\operatorname{\downarrow}}
Neutral and normal forms come equipped with inclusion functions into
ordinary terms and types \(\Tmnf(\Gamma, A) \hookrightarrow \Tm(\Gamma, A)\) etc.
In most cases, the definition for this function will be obvious.
We will also implicitly apply this function when we e.g.\ use a normal term
in a place where a term is expected.

We now begin to give the inductive definition of normal and neutral forms,
transcribing from \bible{\S5.4.1}.
\begin{equation} \tag{\(\star\)} \label{eq:basic-nf}
\inferrule{\Gamma \vdash x : A \isvar}
{\Gamma \vdash x : A \isne} \qquad
\inferrule{\Gamma \vdash A \istype\isne \and
\Gamma \vdash t : A \isne}
{\Gamma \vdash t : A \isnf} \qquad
\inferrule{\Gamma \vdash A \istype\isne}
{\Gamma \vdash A \istype\isnf}
\end{equation}
Strictly speaking, these rules define conversion functions from
variables to neutral forms, and from neutral forms to normal forms, respectively.
If we denote the functions as \(\incl\),
then the conclusions of the inference rules should be written as
\(\Gamma \vdash \incl x : A \isne\) and \(\Gamma \vdash \incl t : A \isnf\), etc.
In most presentations, this conversion is left implicit,
although we shall see that it is not the case for cubical type theories.

\subsection{\texorpdfstring{\(\Pi\)}{Π}-types}
\label{sec:Pi-MLTT}

The rules for \(\Pi\)-types in MLTT reads
\begin{mathpar}
\inferrule{\Gamma \vdash A \istype \and
\Gamma, x \of A \vdash B \istype}
{\Gamma \vdash (x : A) \to B \istype} \\
\inferrule{\Gamma, x \of A \vdash t : B}
{\Gamma \vdash \lambda x \bind t : (x : A) \to B} \and
\inferrule{\Gamma \vdash f : (x : A) \to B \and
\Gamma \vdash t : A}
{\Gamma \vdash f\,t : B[t/x]} \\
\text{(\(\beta, \eta\) rules omitted)}
% \inferrule{\Gamma, x \of A \vdash t : B \and
% \Gamma \vdash s : A}
% {\Gamma \vdash (\lambda x \bind t)s = t[s/x] : B[s/x]} \and
% \inferrule{\Gamma \vdash f : (x : A) \to B}
% {\Gamma \vdash (\lambda x \bind f\,x) = f : (x : A) \to B}
\end{mathpar}
Generally speaking, each introduction rule corresponds to a rule for normal forms,
while each elimination rule corresponds to a rule for neutral forms.
\begin{mathpar}
\inferrule{\Gamma \vdash A \istype \isnf \and
\Gamma, x \of A \vdash B \istype \isnf}
{\Gamma \vdash (x : A) \to B \istype \isnf} \\
\inferrule{\Gamma, x \of A \vdash t : B \isnf}
{\Gamma \vdash \lambda x \bind t : (x : A) \to B \isnf} \and
\inferrule{\Gamma \vdash f : (x : A) \to B \isne \and
\Gamma \vdash t : A \isnf}
{\Gamma \vdash f\,t : B[t/x] \isne}
\end{mathpar}
As a reminder, \(\Gamma \vdash t : A \isnf\) does not presuppose \(A\) is in normal form.
It only requires \(\Gamma \vdash A \istype\) to hold.

\subsection{\texorpdfstring{\(\Sigma\)}{Σ}-types}
The rules for \(\Sigma\)-types in MLTT reads
\begin{mathpar}
\inferrule{\Gamma \vdash A \istype \and
\Gamma, x \of A \vdash B \istype}
{\Gamma \vdash (x : A) \times B \istype} \\
\inferrule{\Gamma \vdash a : A \and
\Gamma \vdash b : B[a/x]}
{\Gamma \vdash (a, b) : (x : A) \times B} \and
\inferrule{\Gamma \vdash p : (x : A) \times B}
{\Gamma \vdash \pi_1(p) : A \and
\Gamma \vdash \pi_2(p) : B[\pi_1(p)/x]} \\
\text{(\(\beta, \eta\) rules omitted)}
\end{mathpar}
To save space, we write multiple conclusions for an inference rule
to mean multiple inference rules with the same premises.
\begin{mathpar}
\inferrule{\Gamma \vdash A \istype\isnf \and
\Gamma, x \of A \vdash B \istype\isnf}
{\Gamma \vdash (x : A) \times B \istype\isnf} \and
\inferrule{\Gamma \vdash a : A \isnf \and
\Gamma \vdash b : B[a/x] \isnf}
{\Gamma \vdash (a, b) : (x : A) \times B \isnf} \and
\inferrule{\Gamma \vdash p : (x : A) \times B \isne}
{\Gamma \vdash \pi_1(p) : A \isne \and
\Gamma \vdash \pi_2(p) : B[\pi_1(p)/x] \isne}
\end{mathpar}

\subsection{Boolean type}
\newcommand{\bool}{\boldsymbol{2}}
\newcommand{\btrue}{\mathrm{true}}
\newcommand{\bfalse}{\mathrm{false}}
\newcommand{\ite}{\mathrm{if}}

The rules for the Boolean type \(\bool\) are given by
\begin{mathpar}
\inferrule{ }{\Gamma \vdash \bool \istype} \and
\inferrule{ }{\Gamma \vdash \btrue : \bool} \and
\inferrule{ }{\Gamma \vdash \bfalse : \bool} \and
\inferrule{\Gamma, x \of \bool \vdash P \istype \and
\Gamma \vdash b : \bool \and
\Gamma \vdash t : P[\btrue/x] \and
\Gamma \vdash f : P[\bfalse/x]}
{\Gamma \vdash \ite_{x \bind P} (b, t, f) : P[b/x]} \and
\text{(\(\beta\) rule omitted)}
\end{mathpar}
which corresponds to the following rules for normal and neutral forms:
\begin{mathpar}
\inferrule{ }{\Gamma \vdash \bool \istype \isnf} \and
\inferrule{ }{\Gamma \vdash \btrue : \bool \isnf} \and
\inferrule{ }{\Gamma \vdash \bfalse : \bool \isnf} \and
\inferrule{\Gamma, x \of \bool \vdash P \istype\isnf \and
\Gamma \vdash b : \bool \isne \\
\Gamma \vdash t : P[\btrue/x] \isnf \and
\Gamma \vdash f : P[\bfalse/x] \isnf}
{\Gamma \vdash \ite_{x \bind P} (b, t, f) : P[b/x] \isne}
\end{mathpar}
However, since the Boolean type doesn't have an \(\eta\) law,
we also include a rule to convert neutral terms into normal terms.
\[\inferrule{\Gamma \vdash b : \bool \isne}{\Gamma \vdash b : \bool \isnf}.\]
Again, strictly speaking the conclusion should read \(\Gamma \vdash \incl b : \bool \isnf\),
but the inclusion function is usually suppressed.

\subsection{Universes}
\newcommand{\univ}{\mathcal{U}}
\newcommand{\El}{\operatorname{El}}
\newcommand{\quo}[1]{\left\lceil #1 \right\rceil}

There are several different presentations for universes.
We will consider Coquand universes and ignore size issues,
but it is straightforward to deal with other formulations.
\begin{mathpar}
\inferrule{ }{\Gamma \vdash \univ \istype} \and
\inferrule{\Gamma \vdash C : \univ}
{\Gamma \vdash \El(C) \istype} \and
\inferrule{\Gamma \vdash A \istype}
{\Gamma \vdash \quo{A} : \univ} \\
\inferrule{\Gamma \vdash C : \univ}
{\Gamma \vdash \quo{\El(C)} = C : \univ} \and
\inferrule{\Gamma \vdash A \istype}
{\Gamma \vdash \El\quo{A} = A \istype}
\end{mathpar}
Here, \(\El\) acts as the eliminator for \(\univ\),
while \(\quo{-}\) acts as the constructor.
In a consistent type theory using Coquand universes,
the type judgments would be indexed by universe levels \(\Gamma \vdash A \istype_i\)
so that not every type has a code in the universe.
\begin{mathpar}
\inferrule{ }{\Gamma \vdash \univ \istype \isnf} \and
\inferrule{\Gamma \vdash C : \univ \isne}
{\Gamma \vdash \El(C) \istype \isne} \and
\inferrule{\Gamma \vdash A \istype\isnf}
{\Gamma \vdash \quo{A} : \univ\isnf}
\end{mathpar}

In a Tarski universe, instead of having a single \(\quo{-}\) constructor,
we have a series of codes for each type connective that the universe is closed under.
There is then a ``$\beta$ equality'' stating that \(\El\) turns these codes
into the actual type connectives,
and there is no $\eta$ equality.
Thus, similar to Boolean types, Tarski universes have the rule
\[\inferrule{\Gamma \vdash C : \univ \isne}{\Gamma \vdash C : \univ \isnf},\]
while Coquand universes do not.

\section{Cubical type theory}

\newcommand{\Path}{\mathrm{Path}}

In general, neutral forms in Martin-L\"of type theory are given by
a stack of eliminators nested on the primary argument,
often referred to as the spine.
The innermost expression is a variable of some type without \(\eta\) laws, such as the Booleans.
Computationally, a neutral form is \emph{stuck} on this variable,
and the computation can only proceed when the value for this variable is known.

The situation is drastically different for cubical type theory.
Here, the path type between two elements acts as a function type from the interval \(\mathbb{I}\),
which has two endpoints \(0, 1 : \mathbb{I}\).
If \(p : \Path_A (x, y)\),
then \(p(0)\) is supposed to be judgmentally equal to \(x\), and \(p(1)\) to \(y\).
Looking at the rules for function types,
we are tempted to say if \(p\) is neutral, then \(p(i)\) should also be neutral.
However, even when \(p\) is a variable, \(p(0)\) still computes to \(x\),
despite being nominally stuck on \(p\)!
See \Cref{rmk:tiny-interval} for why this is a problem in the proof of normalization.

\newcommand{\inst}[1]{{\color[rgb]{0,0.5,0.1}#1}}
For this reason, the definition for neutral forms are greatly complicated.
Instead of defining a set of neutral forms \(\Tmne(\Gamma, A)\) for each \(\Gamma\) and \(A\),
we introduce a third parameter \(\Tmne^{\inst{\varphi}}(\Gamma, A)\),
where \(\inst{\varphi}\) records the condition when the neutral term becomes
unstable and ``decays'' into some other potentially unrelated term.
\(\inst{\varphi}\) is called the \emph{frontier of instability},
and the sequent notation for \(t \in \Tmne^{\inst{\varphi}}(\Gamma, A)\)
is written \(\Gamma \vdash t : A \isne^{\inst{\varphi}}\).

Before we actually define \(\inst{\varphi}\), it should be clear that
the type connectives from Martin-L\"of type theory can simply leave this parameter alone.
For example, the rules for \(\Pi\)-types (\Cref{sec:Pi-MLTT}) become
\begin{mathpar}
\inferrule{\Gamma \vdash A \istype \isnf \and
\Gamma, x \of A \vdash B \istype \isnf}
{\Gamma \vdash (x : A) \to B \istype \isnf} \\
\inferrule{\Gamma, x \of A \vdash t : B \isnf}
{\Gamma \vdash \lambda x \bind t : (x : A) \to B \isnf} \and
\inferrule{\Gamma \vdash f : (x : A) \to B \isne^{\inst{\varphi}} \and
\Gamma \vdash t : A \isnf}
{\Gamma \vdash f\,t : B[t/x] \isne^{\inst{\varphi}}}
\end{mathpar}
Hence most of the rules can be straightforwardly adjusted.
Notable exceptions include rules that turn a neutral form into a normal form,
which we will address later.

\subsection{Interval and cofibrations}

In cubical type theory, the contexts can contain \emph{interval variables}
\(i : \mathbb{I}\), that intuitively represent the coordinates
in an \(n\)-dimensional cube \([0,1]^n\).
However, it should be noted that the interval is not a type,
in the sense that we do not have any element \(\mathbb{I} \in \Tp(\Gamma)\),
and it is not valid to write \(\Tm(\Gamma, \mathbb{I})\).
Instead, we introduce a separate family of sets \(\Tm_{\mathbb{I}}(\Gamma)\)
of interval expressions in a given context,%
\footnote{However, an element \(i \in \Tm_{\mathbb{I}}(\Gamma)\)
is still written as \(\Gamma \vdash i : \mathbb{I}\) in the sequent notation.}
and a separate context extension operation \(\Gamma, i \of \mathbb{I}\).

There are multiple reasons for this distinction.
One is that the interval should not participate in any type formers,
so for example we cannot have functions \(f : \mathbb{N} \to \mathbb{I}\) which
easily make type-checking undecidable.
Another reason is the interval is not Kan,
i.e.\ we cannot define a composition structure (\Cref{sec:composition}) on it,
and it cannot be interpreted semantically as a space.

\begin{remark} \label{rmk:interval-expression}
In Cartesian cubical type theory, the only structures we assume on the interval
is the endpoints \(0, 1 : \mathbb{I}\).
So an expression \(\Gamma \vdash r : \mathbb{I}\)
is essentially either an interval variable \(i \in \Gamma\), or an endpoint \(0, 1\).
The situation is more complicated in de Morgan cubical type theory.
\end{remark}

\newcommand{\iscof}{\;\mathsf{cof}}
\newcommand{\istrue}{\;\mathsf{true}}
To specify portions of a cube, we introduce judgments
\[\Gamma \vdash \varphi \iscof \qquad \Gamma \vdash \varphi \istrue,\]
where the former states that
\(\varphi\) is a formula that is true on the specified portion,
and the latter states the formula is actually true.
We allow contexts to be extended by an assumption of \(\varphi\),
written \(\Gamma, \varphi\).
These formulas are called \emph{cofibrations},
because they semantically represent inclusions \(\Gamma, \varphi \hookrightarrow \Gamma\)
that are intended to act as cofibrations in homotopy theory.

\newcommand{\Cof}{\operatorname{Cof}}
\newcommand{\TrueCof}{\operatorname{True}}
Note that we do not have terms witnessing the proof of \(\Gamma \vdash \varphi \istrue\),
nor do we assign variable names to the assumptions \(\Gamma, \varphi \vdash J\).
The type-checker needs to determine the truth of cofibration formulas automatically.
The cofibrations belong to sets \(\varphi \in \Cof(\Gamma)\),
and we can consider the true cofibrations as a subset.
Alternatively, we can consider a family of sets \(\TrueCof(\Gamma, \varphi)\)
such that every pair of elements \(u, v \in \TrueCof(\Gamma, \varphi)\) are equal.
Then \(\Gamma \vdash \varphi \istrue\) if and only if
\(\TrueCof(\Gamma, \varphi)\) contains a (necessarily unique) element.

In Cartesian cubical type theory, the cofibrations can either be
an atomic cofibration, or finite conjunctions and disjunctions of cofibrations.
An atomic cofibration is given by \((r = s)\) where \(\Gamma \vdash r, s : \mathbb{I}\),
and it is true when \(r\) is in fact equal to \(s\), with equality reflection.
Conjunctions hold if and only if each component holds,
and we have the nullary conjunction \(\Gamma \vdash \top \iscof\).

Similar to the disjoint union type, disjunctions have introduction rules
\begin{mathpar}
\inferrule{\Gamma \vdash \varphi \istrue}
{\Gamma \vdash \varphi \vee \psi \istrue} \and
\inferrule{\Gamma \vdash \psi \istrue}
{\Gamma \vdash \varphi \vee \psi \istrue}
\end{mathpar}
The elimination rule for disjunctions is given by a case split.
Whenever \(\Gamma \vdash \varphi \vee \psi \istrue\)
and some expression is expected, we can instead write
\[\begin{cases*}
  \varphi \hookrightarrow M \\
  \psi \hookrightarrow N
\end{cases*} \quad \text{or} \quad
[
  \varphi \hookrightarrow M;
  \psi \hookrightarrow N
].\]
Naturally, if \(\varphi\) is true, then this expression is equal to \(M\), and similarly for \(\psi\).
Since cofibrations are proof irrelevant, the case split is only allowed
when \(M = N\) judgmentally if \(\varphi\) and \(\psi\) happen to both hold.

This construction is stronger than the rule for the disjoint union type in two ways.
One is that the case split can happen anywhere,
in types, terms, interval expressions, cofibrations,
and (irrelevant) proofs of cofibrations.
Another difference is we require a form of \(\eta\) law to hold for these case splits.
If \(\Gamma, \varphi \vee \psi \vdash t : A\),
then \(t\) is judgmentally equal to a case split
\([\varphi \hookrightarrow t; \psi \hookrightarrow t]\).

\begin{remark}
Many implementations restrict the places a cofibration case split can occur,
to simplify the syntax and type-checking.
We follow the definition of \citet{sterling:2021:thesis},
which includes these case splits, see \bible{6.2\astr14}, \bible{6.4\astr1}.
\end{remark}

For the nullary disjunction \(\bot\), we have no introduction rules,
and the elimination rule is given by an empty case split \([]\),
which means for example \(\Tm((\Gamma, \bot), A)\) always have an element \([]\).
Furthermore, the \(\eta\) law demands that every element is equal to \([]\),
which means every judgmental equality holds when \(\Gamma \vdash \bot \istrue\).

We additionally require cofibrations to satisfy an extensionality rule
\[\inferrule{\Gamma, \varphi \vdash \psi \istrue \and
\Gamma, \psi \vdash \varphi \istrue}
{\Gamma \vdash \varphi = \psi \iscof}\]
So for example \(\varphi \wedge \psi\) is interchangeable with \(\psi \wedge \varphi\),
and \((0 = 1)\) is the same cofibration as \(\bot\).
This implies in particular
the type-checker needs to decide the theory of distributive lattices.

\subsection{Cofibration assumptions}

Since case splits can occur anywhere within a term,
it poses a problem for normal forms.
For example, under the context \(\Gamma, i \of \mathbb{I}, (i = 0) \vee (i = 1)\),
\[f\left(\begin{cases*}
(i = 0) \hookrightarrow s \\
(i = 1) \hookrightarrow t
\end{cases*}\right) \quad \text{and} \quad
\begin{cases*}
(i = 0) \hookrightarrow f(s) \\
(i = 1) \hookrightarrow f(t)
\end{cases*}\]
are judgmentally equal, if \(f\) is a function and \(s, t\) are terms.
To normalize this, we require case splits to always occur \emph{up front} \bible{7.1\astr4}.
More generally, when a cofibration is about to be introduced into the context,
we require the normal form to completely decompose the cofibration,
so whenever we write \(\Gamma \vdash t : A \isnf\) we can safely assume
\(\Gamma\) does not actually contain any cofibration.

\newcommand{\dom}{\operatorname{dom}}
Concretely, we recursively define sets \([\varphi]\,\Tmnf(\Gamma, A)\),
whose elements \(t \in [\varphi]\,\Tmnf(\Gamma, A)\) will be written as
\(\Gamma, [\varphi] \vdash t : A \isnf\).
In addition, we can convert \(t\) into a term \(\Gamma, \varphi \vdash \incl t : A\).
We will still suppress this conversion function, unless it causes confusion.
An element \(t \in [\varphi]\,\Tmnf(\Gamma, A)\) is required to be a function
\(\varphi' \mapsto t_{\varphi'}\) whose domain is a subset of \(\Cof(\Gamma)\),
satisfying the following conditions.
\begin{itemize}
\item \(t \in [\top]\,\Tmnf(\Gamma, A)\) iff its domain \(\{\top\}\), and \(t_{\top} \in \Tmnf(\Gamma, A)\).
\item \(t \in [\varphi \wedge \psi]\,\Tmnf(\Gamma, A)\) iff
there is a binary function \(t' \in [\varphi][\psi]\,\Tmnf(\Gamma, A)\),
such that \(t_{\varphi' \wedge \psi'}\) is set to \((t'{}_{\varphi'})_{\psi'}\),
whenever this is in the domain of \(t'\).
\item \(t \in [\bot]\,\Tmnf(\Gamma, A)\) iff its domain is empty.
\item \(t \in [\varphi \vee \psi]\,\Tmnf(\Gamma, A)\) iff it is the union%
\footnote{Recall the union of two functions \(f \cup g\) is defined
when they agree on the intersection of their domains, and
it maps \(x\) to either \(f(x)\) or \(g(x)\), whichever is defined.}
of two functions belonging to \([\varphi]\,\Tmnf(\Gamma, A)\)
and \([\psi]\,\Tmnf(\Gamma, A)\), respectively.
\item \(t \in [r = s]\,\Tmnf(\Gamma, A)\) iff,
recalling \Cref{rmk:interval-expression}, one of the following holds:
\begin{itemize}
\item The equation is \(0 = 1\), and the domain is empty.
\item The equation is \(r = r\), \(\dom(t) = \{r = r\}\)
and \(t_{r = r} \in \Tmnf(\Gamma, A)\).
\item The equation is \(i = r\) where \(i\) is a variable,
the domain is \(\{i = r\}\), and \(t_{i = r} \in \Tmnf(\Gamma', A[r/i])\),
where \(\Gamma'\) is the context \(\Gamma\) removing \(i\),
while substituting all its occurrences with \(r\).
\end{itemize}
\end{itemize}
For example, a normal form in the context \(\Gamma, i \of \mathbb{I}, j \of \mathbb{I}, (i = 0) \vee (i = j)\)
is of the form
\[
\Gamma, i \of \mathbb{I}, j \of \mathbb{I} \vdash \begin{cases*}
  (i = 0) \hookrightarrow t_1 \\
  (i = j) \hookrightarrow t_2
\end{cases*} : A \isnf
\]
where \(\Gamma, j : \mathbb{I} \vdash t_1 : A[0/i] \isnf\) and
\(\Gamma, i : \mathbb{I} \vdash t_2 : A[i/j] \isnf\).

For this to be well-defined, we need to verify if \(\varphi = \varphi'\),
then \([\varphi]\,\Tmnf(\Gamma, A)\) is \emph{equal} as a set to \([\varphi']\,\Tmnf(\Gamma, A)\).
This is left as an exercise.
Identical constructions can be carried out for normal types, and neutral terms and types.

\subsection{Frontier of instability}

We are now ready to define the frontier of instability \(\inst{\varphi}\),
which is simply a cofibration in the current context.
So \(\Tmne^{\inst{\varphi}} (\Gamma, A)\) presupposes that
\(\Gamma \in \Ctx\), \(A \in \Tp(\Gamma)\) and \(\varphi \in \Cof(\Gamma)\),
and similarly for the other sets.
For the simplest case, variables are never unstable \bible{7.2.1\astr2}:
\begin{mathpar}
\inferrule{\Gamma \vdash x : A \isvar}
{\Gamma \vdash x : A \isne^{\inst{\bot}}}
\end{mathpar}

Next, we consider the rules for non-dependent path types:
\begin{mathpar}
\inferrule{\Gamma \vdash A \istype \and
\Gamma \vdash t, s : A}
{\Gamma \vdash \Path_A (t, s) \istype} \and
\inferrule{\Gamma, i \of \mathbb I \vdash a : A \and
\Gamma \vdash a[0/i] = t : A \and
\Gamma \vdash a[1/i] = s : A}
{\Gamma \vdash (\lambda i \bind a) : \Path_A (t, s)} \and
\inferrule{\Gamma \vdash p : \Path_A (t, s) \and
\Gamma \vdash r : \mathbb{I}}
{\Gamma \vdash p(r) : A \and
\Gamma \vdash p(0) = t : A \and
\Gamma \vdash p(1) = s : A} \and
\text{($\beta, \eta$ laws omitted)}
\end{mathpar}
Note the judgmental equalities \emph{in addition to}
the ordinary \(\beta\) and \(\eta\) laws.

A neutral term \(p(i)\) will reduce when \(i\) is \(0\) or \(1\),
so we record this in the frontier of instability, following \bible{7.2\astr9}.
\begin{mathpar}
\inferrule{\Gamma \vdash A \istype\isnf \and
\Gamma \vdash t, s : A \isnf}
{\Gamma \vdash \Path_A (t, s) \istype\isnf} \and
\inferrule{\Gamma, i \of \mathbb I \vdash a : A \isnf \and
\Gamma \vdash a[0/i] = t : A \and
\Gamma \vdash a[1/i] = s : A}
{\Gamma \vdash (\lambda i \bind a) : \Path_A (t, s) \isnf} \and
\inferrule{\Gamma \vdash p : \Path_A (t, s) \isne^{\inst{\varphi}} \and
\Gamma \vdash r : \mathbb{I}}
{\Gamma \vdash p(r) : A \isne^{\inst{\varphi \vee (r = 0) \vee (r = 1)}}}
\end{mathpar}

\begin{remark}
Note that \(r : \mathbb{I}\) here is not required to be in normal form.
Instead, we directly use the interval expressions up to judgmental equality
in the definition of normal and neutral forms.
This is fine, because the judgmental equality for
interval expressions and cofibrations is decidable.
\end{remark}

Since neutral forms may decay,
we cannot simply include them in normal forms.
Instead, we need to provide a \emph{stabilization} \bible{7.2\astr2},
i.e.\ a normal form that the destabilized neutral form reduces to.
Consider the Booleans:
\[\inferrule{\Gamma \vdash b : \bool \isne^{\inst{\varphi}} \and
\Gamma, [\varphi] \vdash \tilde{b} : \bool \isnf \and
\Gamma, \varphi \vdash b = \tilde{b} : \bool}
{\Gamma \vdash \incl^{\inst{\varphi}}(b, \tilde{b}) : \bool \isnf}.\]
Intuitively, the neutral form \(b\) can be taken as the normal form
as long as \(\varphi\) is not true;
when it becomes true, we switch to using the backup normal form \(\tilde{b}\).

Since we expect judgmentally equal terms to have equal normal forms,
when a neutral destabilizes, we need to remove the redundant information:
\begin{mathpar}
\inferrule{\Gamma \vdash \varphi \istrue \\
\Gamma \vdash t : A}
{\Gamma \vdash \coll t : A \isne^{\inst{\varphi}}} \and
\inferrule{\Gamma \vdash a : A \isne^{\inst{\varphi}} \and
\Gamma \vdash \varphi}
{\Gamma \vdash a = \mathop{\downarrow\uparrow} a : A \isne^{\inst{\varphi}}} \and
\inferrule{\Gamma \vdash \varphi \istrue \\
\Gamma \vdash b : \bool \isne^{\inst{\varphi}} \and
\Gamma, [\varphi] \vdash \tilde{b} : \bool \isnf \and
\Gamma, \varphi \vdash b = \tilde{b} : \bool}
{\Gamma \vdash \incl^{\inst{\varphi}}(b, \tilde{b}) = \tilde{b} : \bool \isnf}
\end{mathpar}
Here, $\mathop{\downarrow\uparrow} a$ first coerces a neutral form into
a term, and then uses the $\coll$ constructor to turn it into
a neutral form again. These should be seen as constructors in the quotient-inductive
construction of the sets \(\Tmne^{\inst{\varphi}}(\Gamma, A)\) and \(\Tmnf(\Gamma, A)\) \bible{7.2.1\astr1}.
When \(\varphi\) is true, \(\Tmne^{\inst{\varphi}}(\Gamma, A)\) becomes
in bijection to the set $\Tm(\Gamma, A)$,\footnote{Since we always need a well-defined map
\(\Tmne^{\inst{\varphi}}(\Gamma, A) \to \Tm(\Gamma, A)\), collapsing the two to
be in bijection erases all the information we can erase.}
and the stabilized normal form \(\incl^{\inst{\varphi}} (..., \tilde{b})\)
simply becomes the provided backup \(\tilde{b}\).
We have similar rules for neutral types.

Recall the neutral conversion rule for neutral types in \Cref{eq:basic-nf}.
It is also upgraded to use stabilized neutrals \bible{7.2.1\astr2}:
\begin{mathpar}
\inferrule{\Gamma \vdash A \istype\isne^{\inst{\varphi}} \and
\Gamma \vdash t : A \isne^{\inst{\psi}} \and
\Gamma, [\varphi \vee \psi] \vdash \tilde{t} : A \isnf \\
\Gamma, \varphi \vee \psi \vdash t = \tilde{t} : A}
{\Gamma \vdash \incl^{\inst{\varphi}, \inst{\psi}}(A, t, \tilde{t}) : A \isnf} \and
\inferrule{\Gamma \vdash A \istype\isne^{\inst{\varphi}} \and
\Gamma, [\varphi] \vdash \tilde{A} \istype\isnf \and
\Gamma, \varphi \vdash A = \tilde{A} \istype}
{\Gamma \vdash \incl^{\inst{\varphi}}(A, \tilde{A}) \istype\isnf}
\end{mathpar}

\begin{remark} \label{rmk:tiny-interval}
All of these constructions are effectively used to guarantee that,
given a normal form \(t\) with some interval variable \(i\),
we can straightforwardly compute the normal form of e.g.\ the substitution \(t[0/i]\).
However, why go through all the trouble if we can simply ban these substitutions?
Indeed, in the proof of normalization for Martin-L\"of type theory,
we consider computability structures over the category of \emph{renamings} only,
and normal forms are preserved under renamings.

A reason specific to Cartesian cubical type theory is
that contracting two variables \(t[k/i, k/j]\) also causes the neutral to destabilize.
A deeper and more essential reason is that
normal forms involving composition structures can still compute,
which requires arbitrary substitutions on interval variables.
% In category theoretic language, the normalization proof requires
% the existence of an exceptional right adjoint \((-)^{\mathbb{I}} \dashv \sqrt[\mathbb{I}]{-}\)
% in the category of computability structures.
See \bible{7.3.1\astr5} and \citet{licata:2018:universes}.
\end{remark}

For completeness, we include the rules for dependent path types,
which are given by simple modifications of the non-dependent version.
\begin{mathpar}
\inferrule{\Gamma, i \of \mathbb{I} \vdash A \istype \and
\Gamma \vdash t : A[0/i] \and
\Gamma \vdash s : A[1/i]}
{\Gamma \vdash \Path_{i\bind A} (t, s) \istype} \and
\inferrule{\Gamma, i \of \mathbb I \vdash a : A \and
\Gamma \vdash a[0/i] = t : A[0/i] \and
\Gamma \vdash a[1/i] = s : A[1/i]}
{\Gamma \vdash (\lambda i \bind a) : \Path_A (t, s)} \and
\inferrule{\Gamma \vdash p : \Path_{i \bind A} (t, s) \and
\Gamma \vdash r : \mathbb{I}}
{\Gamma \vdash p(r) : A[r/i] \and
\Gamma \vdash p(0) = t : A[0/i] \and
\Gamma \vdash p(1) = s : A[1/i]}
\end{mathpar}
The normal and neutral forms are then given by
\begin{mathpar}
\inferrule{\Gamma, i \of \mathbb{I} \vdash A \istype\isnf \and
\Gamma \vdash t : A[0/i] \isnf \and
\Gamma \vdash s : A[1/i] \isnf}
{\Gamma \vdash \Path_{i\bind A} (t, s) \istype\isnf} \and
\inferrule{\Gamma, i \of \mathbb I \vdash a : A \isnf \and
\Gamma \vdash a[0/i] = t : A[0/i] \and
\Gamma \vdash a[1/i] = s : A[1/i]}
{\Gamma \vdash (\lambda i \bind a) : \Path_A (t, s) \isnf} \and
\inferrule{\Gamma \vdash p : \Path_{i \bind A} (t, s) \isne^{\inst{\varphi}} \and
\Gamma \vdash r : \mathbb{I}}
{\Gamma \vdash p(r) : A[r/i] \isne^{\inst{\varphi \vee (r = 0) \vee (r = 1)}}}
\end{mathpar}

\subsection{Composition structure} \label{sec:composition}

\newcommand{\hcomp}{\operatorname{hcomp}}
\newcommand{\comp}{\operatorname{comp}}
\newcommand{\coe}{\operatorname{coe}}
We turn to composition structures, a.k.a.\ the Kan structures on each type.
For Cartesian cubical type theory, this is often decomposed into
a homogeneous composition and a coercion structure.
The former is given by
\begin{mathpar}
\inferrule{\Gamma \vdash A \istype \and
\Gamma \vdash r, s : \mathbb{I} \and
\Gamma \vdash \varphi \iscof \and
\Gamma, i \of \mathbb{I}, (i = r) \vee \varphi \vdash t : A}
{\Gamma \vdash \hcomp_{r \rightsquigarrow s}^{\varphi}(i \bind t) : A} \and
\inferrule{\text{(Same premises)} \and
\Gamma \vdash (r = s) \vee \varphi \istrue}
{\Gamma \vdash \hcomp_{r \rightsquigarrow s}^{\varphi}(i \bind t) = t[s/i] : A}
\end{mathpar}
and the latter has rules
\begin{mathpar}
\inferrule{\Gamma, i \of \mathbb{I} \vdash A \istype \and
\Gamma \vdash r, s : \mathbb{I} \and
\Gamma \vdash t : A[r/i]}
{\Gamma \vdash \coe_{r \rightsquigarrow s}(i \bind A, t) : A[s/i]} \and
\inferrule{\text{(Same premises with \(s\) removed)}}
{\Gamma \vdash \coe_{r \rightsquigarrow r}(i \bind A, t) = t : A[r/i]}
\end{mathpar}
We can combine them to obtain a heterogeneous composition
\begin{mathpar}
\inferrule{\Gamma, i \of \mathbb{I} \vdash A \istype \and
\Gamma \vdash r, s : \mathbb{I} \and
\Gamma \vdash \varphi \iscof \and
\Gamma, i \of \mathbb{I}, (i = r) \vee \varphi \vdash t : A}
{\Gamma \vdash \comp_{r \rightsquigarrow s}^{\varphi}(i \bind A, i \bind t) : A[s/i]} \and
\comp^{\varphi}_{r \rightsquigarrow s}(i \bind A, i \bind t) \coloneq
\hcomp_{r \rightsquigarrow s}^{\varphi} (i \bind \coe_{i \rightsquigarrow s} (i \bind A, t))
\end{mathpar}
They also have computation rules when \(A\) is given by concrete type connectives.
We will not write them down here.
If \(A\) is a neutral type,
the composition structure is stuck and produces a normal form \bible{7.2\astr7}:
\begin{mathpar}
\inferrule{\Gamma \vdash A \istype\isne^{\inst{\psi}} \and
\Gamma \vdash r, s : \mathbb{I} \and
\Gamma \vdash \varphi \iscof \\
\Gamma, i \of \mathbb{I}, [(i = r) \vee \varphi] \vdash t : A \isnf \\
\Gamma, [\psi] \vdash \tilde{h} : A \isnf \\
\Gamma, \psi \vdash \tilde{h} = \hcomp_{r \rightsquigarrow s}^{\varphi}(i \bind t) : A \and
\Gamma, [\psi], [(r = s) \vee \varphi] \vdash \tilde{h} = t[s/i] : A \isnf}
{\Gamma \vdash \hcomp_{r \rightsquigarrow s}^{\varphi, \inst{\psi}}(i \bind t, \tilde{h}) : A \isnf} \and
\inferrule{\Gamma, i \of \mathbb{I} \vdash \varphi \iscof \and
\Gamma, i \of \mathbb{I} \vdash A \istype\isne^{\inst{\varphi}} \and
\Gamma \vdash r, s : \mathbb{I} \and
\Gamma \vdash t : A[r/i] \isnf \\
\Gamma, [\forall i\bind \varphi] \vdash \tilde{c} : A[s/i] \isnf \\
\Gamma, (\forall i\bind \varphi) \vdash \tilde{c} = \coe_{r \rightsquigarrow s}(i \bind A, t) : A[s/i] \\
\Gamma, [\forall i\bind \varphi], [r = s] \vdash \tilde{c} = t : A[r/i] \isnf}
{\Gamma \vdash \coe_{r \rightsquigarrow s}^{\inst{\varphi}}(i \bind A, t, \tilde{c}) : A[s/i] \isnf}
\end{mathpar}
A few things to note here. Firstly, they both need stabilization structures.
Secondly, most frontiers of instability are in the obvious context \(\Gamma\),
but for coercions, the frontier \(\varphi\) also depends on the new interval variable \(i\),
so the premise for \(\varphi\) is explicitly written down.
Thirdly, the cofibration \((\forall i\bind \varphi)\) appeared in the context.
This is a cofibration that holds if and only if \(\varphi\) holds regardless of \(i\).

The universal quantification on cofibrations is required for 
the composition structure on glue types (see \citet{orton:2018:axioms}),
and so it's reasonable to see it in the composition structure on neutral types here.
Fortunately, cubical type theories usually enjoy
a form of \emph{quantification elimination} \cite{cchm:2018:cubical},
so \((\forall i \bind \varphi)\) is equivalent to a cofibration not involving \(\forall\).
\begin{itemize}
\item The cofibration \(\forall i \bind \varphi \wedge \psi\) is
always equivalent to \((\forall i \bind \varphi) \wedge (\forall i \bind \psi)\)
by the inference rules.
\item If \(\forall i \bind \varphi \vee \psi\) is true,
then \((\forall i \bind \varphi) \vee (\forall i \bind \psi)\) holds,
because it is impossible to cover an entire interval using two cofibrations,
unless one of them already covers the interval.
\item If \(\forall i \bind (r = s)\) is true,
we can remove the quantification if \(r, s\) don't involve \(i\),
and otherwise the cofibration must be equivalent to \(\top\) or \(\bot\).
\end{itemize}
Hence, we can regard this operation as a syntactic transformation
on the meta-language level.

We add the equations for destabilization as follows.
For \(\hcomp\), if \(\psi\) holds in \(\Gamma\), then
the normal form \(\Gamma \vdash \hcomp^{\varphi, \inst{\psi}}_{r \rightsquigarrow s} (i \bind t, \tilde{h})\)
is equal to the stabilizer \(\tilde{h}\).
If \((r = s) \vee \varphi\) is true, then the normal form is equal to \(t[s/i]\).
Similarly for \(\coe\), if \(\forall i \bind \psi\) holds,
then \(\Gamma \vdash \coe_{r \rightsquigarrow s}^{\inst{\varphi}}(i \bind A, t, \tilde{c}) = \tilde{c} : A[s/i] \isnf\).
If \((r = s)\) is true, then it is equal to \(t\).

\subsection{Weak Booleans} \label{sec:weak-boolean}

Ordinary inductive types bifurcate into two variants in cubical type theory,
based on how \(\hcomp\) interacts with them.
One version has \(\hcomp\) compute on constructors,
schematically written as
\begin{equation} \tag{\textdagger} \label{eq:hcomp-strict}
\hcomp^{\varphi}_{r \rightsquigarrow s} (i \bind \mathrm{cons}(t))
= \mathrm{cons}(\hcomp^{\varphi}_{r \rightsquigarrow s} (i \bind t))
\end{equation}
This is valid because for inductive types,
different constructors are distinct,
and constructors are injective functions (or embeddings, homotopically speaking).
Therefore if argument of \(\hcomp\) ---
defined in the context of \(\varphi\), i.e.\ a portion of a cube ---
evaluates to a constructor at any point in the cube,
then it has to evaluate to the same constructor everywhere,
and we can uniquely recover the arguments of the constructor.

The other version comes from considering inductive types as a special case
of higher inductive types, and directly copying their rules.
Crucially, it is possible in a HIT for an expression \(t(i)\) to be
a constructor when \(i = 0\), and a different one when \(i = 1\).
\Cref{eq:hcomp-strict} is therefore no longer adequate.
Instead, we consider \(\hcomp\) over higher inductive types as new values,
and eliminators for HITs define computation rules over \(\hcomp\).
In this case, we need to add normal \(\hcomp\) terms,
following \bible{7.2\astr4} and \bible{7.2\astr11}.
\begin{mathpar}
\inferrule{\Gamma \vdash \varphi \iscof \and
\Gamma \vdash r, s : \mathbb{I} \and
\Gamma, i \of \mathbb{I}, [(i = r) \vee \varphi] \vdash b : \bool \isnf}
{\Gamma \vdash \hcomp^{\varphi}_{r \rightsquigarrow s} (i \bind b) : \bool \isnf} \and
\inferrule{\text{(Same premises)} \and
\Gamma \vdash (r = s) \vee \varphi}
{\Gamma \vdash \hcomp^{\varphi}_{r \rightsquigarrow s} (i \bind b) = b[s/i] : \bool \isnf}
\end{mathpar}

\subsection{The circle}

With the warm-up from \Cref{sec:weak-boolean},
we are ready to consider the circle \(\mathbb{S}^1\).
\newcommand{\Sbase}{\mathrm{base}}
\newcommand{\Sloop}{\mathrm{loop}}
\newcommand{\Selim}{\operatorname{elim}}
\begin{mathpar}
\inferrule{ }{\Gamma \vdash \mathbb{S}^1 \istype} \and
\inferrule{ }{\Gamma \vdash \Sbase : \mathbb{S}^1} \and
\inferrule{\Gamma \vdash r : \mathbb{I}}
{\Gamma \vdash \Sloop_r : \mathbb{S}^1} \and
\inferrule{ }{\Gamma \vdash \Sloop_0 = \Sloop_1 = \Sbase : \mathbb{S}^1} \and
\inferrule{\Gamma, x \of \mathbb{S}^1 \vdash P \istype \and
\Gamma \vdash t : \mathbb{S}^1 \\
\Gamma \vdash b : P[\Sbase/x] \and
\Gamma, i \of \mathbb{I} \vdash \ell : P[\Sloop_i/x] \\
\Gamma \vdash \ell[0/i] = \ell[1/i] = b : P[\Sbase/x]}
{\Gamma \vdash \Selim_{x \bind P} (t, b, i \bind \ell) : P(t)}
\end{mathpar}
The \(\beta\) equation for eliminators will include a clause
where \(t\) is given by some \(\hcomp\) over \(\mathbb{S}^1\).

Following \bible{7.2\astr12}, the normal and neutral forms are given by
\begin{mathpar}
\inferrule{ }{\Gamma \vdash \mathbb{S}^1 \istype\isnf} \and
\inferrule{ }{\Gamma \vdash \Sbase : \mathbb{S}^1 \isnf} \and
\inferrule{\Gamma \vdash r : \mathbb{I}}
{\Gamma \vdash \Sloop_r : \mathbb{S}^1 \isnf} \and
\inferrule{ }{\Gamma \vdash \Sloop_0 = \Sloop_1 = \Sbase : \mathbb{S}^1 \isnf} \and
\inferrule{\Gamma, x \of \mathbb{S}^1 \vdash P \istype \isnf \and
\Gamma \vdash t : \mathbb{S}^1 \isne^{\inst{\varphi}} \\
\Gamma \vdash b : P[\Sbase/x] \isnf \and
\Gamma, i \of \mathbb{I} \vdash \ell : P[\Sloop_i/x] \isnf \\
\Gamma \vdash \ell[0/i] = \ell[1/i] = b : P[\Sbase/x] \isnf}
{\Gamma \vdash \Selim_{x \bind P} (t, b, i \bind \ell) : P(t) \isne^{\inst{\varphi}}}
\end{mathpar}
And as expected there are normal \(\hcomp\)s.
The rules are identical to that of the weak Booleans,
with only the types changed.
\begin{mathpar}
\inferrule{\Gamma \vdash \varphi \iscof \and
\Gamma \vdash r, s : \mathbb{I} \and
\Gamma, i \of \mathbb{I}, [(i = r) \vee \varphi] \vdash t : \mathbb{S}^1 \isnf}
{\Gamma \vdash \hcomp^{\varphi}_{r \rightsquigarrow s} (i \bind t) : \mathbb{S}^1 \isnf} \and
\inferrule{\text{(Same premises)} \and
\Gamma \vdash (r = s) \vee \varphi}
{\Gamma \vdash \hcomp^{\varphi}_{r \rightsquigarrow s} (i \bind t) = t[s/i] : \mathbb{S}^1 \isnf}
\end{mathpar}

\subsection{Glue types} \label{sec:glue}

\newcommand{\Glue}{\operatorname{Glue}}
\newcommand{\glue}{\operatorname{glue}}
\newcommand{\unglue}{\operatorname{unglue}}
\(\Glue\) has some of the most complex rules in cubical type theory.
We write \(e : A \simeq B\) for an equivalence between \(A\) and \(B\),
and abuse notation to use \(e\) as a function.
Usually we define equivalence as having contractible fibers,
but for the purpose of defining normal forms the concrete definition does not matter.

The parameters for the glue type are complicated enough that
we give a name for them.
Call the following a set of \textbf{gluing data}:
\[\Gamma \vdash \varphi \iscof \qquad
\Gamma \vdash B \istype \qquad
\Gamma, \varphi \vdash A \istype \qquad
\Gamma, \varphi \vdash e : A \simeq B.\]
Naturally, we can define \emph{normal} gluing data as
\[\Gamma \vdash \varphi \iscof \qquad
\Gamma \vdash B \istype \isnf \qquad
\Gamma, [\varphi] \vdash A \istype \isnf \qquad
\Gamma, [\varphi] \vdash e : A \simeq B \isnf.\]

The basic rules for the \(\Glue\) type is given by
\begin{mathpar}
\inferrule{\Gamma \vdash \text{(Gluing data)}}
{\Gamma \vdash \Glue^{\varphi} (B, A, e) \istype} \and
\inferrule{\Gamma \vdash \text{(Gluing data)}\and
\Gamma \vdash \varphi \istrue}
{\Gamma \vdash \Glue^{\varphi} (B, A, e) = A \istype} \and
\inferrule{\Gamma \vdash b : B \and
\Gamma, \varphi \vdash a : A \and
\Gamma, \varphi \vdash b = e(a) : B}
{\Gamma \vdash \glue^{\varphi}(b, a) : \Glue^{\varphi}(B, A, e)} \and
\inferrule{\text{(Same premises)} \and
\Gamma \vdash \varphi \istrue}
{\Gamma \vdash \glue^{\varphi}(b, a) = a : A} \\
\inferrule{\Gamma \vdash g : \Glue^{\varphi} (B, A, e)}
{\Gamma \vdash \unglue(g) : B} \and
\inferrule{\Gamma \vdash g : \Glue^{\varphi} (B, A, e) \and
\Gamma \vdash \varphi \istrue}
{\Gamma \vdash \unglue(g) = e(g) : B}
\end{mathpar}
The normal and neutral forms are given in~\bible{7.1\astr10}.
Naturally, \(\varphi\) enters the frontier of instability.
\begin{mathpar}
\inferrule{\Gamma \vdash \text{(Gluing data)} \isnf}
{\Gamma \vdash \Glue^{\varphi} (B, A, e) \istype\isnf} \and
\inferrule{\Gamma \vdash \text{(Gluing data)} \isnf \and
\Gamma \vdash \varphi \istrue}
{\Gamma \vdash \Glue^{\varphi} (B, A, e) = A \istype\isnf} \and
\inferrule{\Gamma \vdash b : B \isnf \and
\Gamma, [\varphi] \vdash a : A \isnf \and
\Gamma, \varphi \vdash b = e(a) : B}
{\Gamma \vdash \glue^{\varphi}(b, a) : \Glue^{\varphi}(B, A, e) \isnf} \and
\inferrule{\text{(Same premises)} \and
\Gamma \vdash \varphi \istrue}
{\Gamma \vdash \glue^{\varphi}(b, a) = a : A \isnf} \and
\inferrule{\Gamma \vdash g : \Glue^{\varphi} (B, A, e) \isne^{\inst{\psi}}}
{\Gamma \vdash \unglue(g) : B \isne^{\inst{\psi \vee \varphi}}}
\end{mathpar}

\section{Equality on normal forms}

There are many equality constructors in the quotient-inductive definition
of cubical normal forms,
so it is not entirely obvious that equality between normal forms are decidable.
Here we give a rough justification.
The equalities in normal forms fall into three classes:
\begin{itemize}
\item For cleaning up destabilized neutrals, including stuck \(\hcomp\) and \(\coe\);
\item For \(\Glue\) and \(\hcomp\) terms on higher inductive types;
\item For the boundary of path constructors such as \(\Sloop_r\).
\end{itemize}
However, we have carefully constructed these equations so there is
always a clear direction in which the complexity of (raw, unquotiented) terms decreases.
Therefore, it suffices to recursively rewrite using these equations.

The other source of non-trivial equalities in normal forms
is indirectly through interval and cofibration sub-expressions.
These are decidable using the theory of distributive lattices.

\begin{remark}
Deciding equality for normal forms has mostly theoretical value,
since practical implementations of type-checkers check for
judgmental equality directly on values,
which are essentially given by the computability structures in a normalization proof.
\end{remark}

\appendix
\section{Identity types and complex inductive types}

A major extension to our cubical type theory is to add more flexible forms
of higher inductive types, which require some care
to define the correct composition and coercion rules.
In particular, this also includes parametrized and indexed inductive types
without higher constructors.

\newcommand{\List}{\operatorname{List}}
\newcommand{\Susp}{\operatorname{Susp}}
\newcommand{\Id}{\operatorname{Id}}
\newcommand{\refl}{\mathrm{refl}}
For (higher) inductive types with parameters such as lists \(\List(A)\) and suspensions \(\Susp(A)\),
the non-trivial work is to define \(\coe_{r \rightsquigarrow s}\) on the type \(i \bind \List(A(i))\)
in terms of the coercion on the type \(i \bind A(i)\).
However, the normal forms remain the same, so we will not further elaborate.
The reader can consult, e.g.~\citet[\S 2.15]{abcfhl:2021:cubical} for the rules.

The case of indexed inductive types (a.k.a.~\emph{inductive families}) is much trickier,
as coercions no longer have a suitable value to compute to,
and must also introduce new normal forms much like \(\hcomp\) in higher inductive types.

As an example, the prototypical indexed inductive type is the Martin-L\"of identity type
\(\Id_A (t, t')\), with a constructor \(\refl_t : \Id_A (t, t)\).
Suppose we have a path \(p : \Path_A (t, t')\),
we could use coercion on \(\refl_{t'}\) along the type \(i \bind \Id_A (p(i), t')\)
to obtain an element of type \(\Id_A (t, t')\).
This cannot be \(\refl\) since \(t\) and \(t'\) may not be judgmentally equal.
This suggests that the identity type should somehow ``contain'' the same content
as the path type. Indeed, for \emph{any} type family \(\Id_A (t, t')\) with an element \(\refl_t\),
there is a map \(\Path_A (t, t') \to \Id_A (t, t')\), given by
\[p \mapsto \coe_{0 \rightsquigarrow 1} (i. \Id_A (t, p(i)), \refl_t).\]

However, the cubical path type cannot be directly used as a substitute
for the identity type, because the judgmental equality
\[\tag{\S} \label{eq:J} J_{xp \bind P}(t, \refl_t, r) = r[t/x]\]
cannot be made true for path types.
On the surface level, this is because \(\hcomp\) computes by rules
specific to each type constructor, and so it does not simplify on
the constant path \(\refl_t = \lambda i \bind t\) unless the type of \(t\) is known.
See \citet{swan:2018:path} for some negative results in this direction.

\newcommand{\trX}{\operatorname{cmp}}
On the other hand, a cubical adaptation of the identity type was developed early on,
known as the \emph{Swan identity type}.
The basic idea is to define a type with the constructor
\(\trX^{\varphi}(i \bind p) : \Id_A (t, t')\) where
\(p\) is a path from \(t\) to \(t'\).
The crucial ingredient is the extra cofibration \(\varphi\),
recording whether \(p\) is the constant path.
\[
\inferrule{\Gamma \vdash \varphi \iscof \and
\Gamma, i \of \mathbb{I} \vdash p : A \and
\Gamma, \varphi, i \of \mathbb{I} \vdash p[0/i] = p = p[1/i] : A}
{\Gamma \vdash \trX^{\varphi}(i \bind p) : \Id_A (p[0/i], p[1/i])}
\]
The reflexivity element of \(\Id_A (t, t)\) is defined as
\(\refl_t \coloneq \trX^\top (i \bind t)\),
with the cofibration signifying that it is always constant.
Our goal is now to define a $J$ operator that computes on \(\trX^{\varphi}(i \bind p)\)
to some cubical expression involving \(\varphi\) and \(p\), so that
it satisfies the same judgmental equalities that we would expect in Martin-L\"of type theory.
This is done by \citet[\S 2.16]{abcfhl:2021:cubical},
and the de Morgan variant is described by \citet{cchm:2018:cubical}.
The idea is to use the fact that \(\hcomp^{\varphi}_{r \rightsquigarrow s}(i \bind t)\)
is judgmentally equal to \(t[s/i]\) when \(\varphi\) is actually true,
which grants us the ability to make \Cref{eq:J} judgmentally true.

Under this mechanism, we can extract the rules for normal forms straightforwardly.
If \(p\) is a normal form, then \(\trX^{\varphi} (i \bind p)\) is also a normal form.
Since identity types have no \(\eta\) laws, neutral terms of this type can be normal.
The $J$ operator can get stuck on a neutral term.
\begin{mathpar}
\inferrule{\Gamma \vdash \varphi \iscof \and
\Gamma, i \of \mathbb{I} \vdash p : A \isnf \and
\Gamma, \varphi, i \of \mathbb{I} \vdash p[0/i] = p = p[1/i] : A}
{\Gamma \vdash \trX^{\varphi}(i \bind p) : \Id_A (p[0/i], p[1/i]) \isnf} \and
\inferrule{\Gamma \vdash q : \Id_A (t, t') \isne^{\inst{\varphi}} \and
\Gamma, [\varphi] \vdash \tilde{q} : \Id_A (t, t') \isnf \and
\Gamma, \varphi \vdash q = \tilde{q} : \Id_A (t, t')}
{\Gamma \vdash \incl^{\inst{\varphi}}(q, \tilde{q}) : \Id_A (t, t') \isnf} \and
\inferrule{\Gamma \vdash t' : A \isnf \\
\Gamma \vdash q : \Id_A (t, t') \isne^{\inst{\varphi}} \\
\Gamma \vdash r : P[t/x, \refl_t/q] \isnf}
{\Gamma \vdash J_{xq \bind P} (t', q, r) \isne^{\inst{\varphi}}}
\end{mathpar}
We can see the rules are essentially analogous to the Boolean case,
and the complexity of the Swan identity type lies in the computation rules.

Intuitively, the constructor \(\trX\) can be understood as a formal composition
\begin{align*}
\trX^{\varphi}(i \bind p) &\approx
\comp^{\varphi}_{0 \rightsquigarrow 1}(i \bind \Id_A (t, p), i \bind \refl_t) \\
&\coloneq
\hcomp_{0 \rightsquigarrow 1}^{\varphi} (i \bind \coe_{i \rightsquigarrow 1} (i \bind \Id_A (t, p), \refl_t))
\end{align*}
where the endpoint \(p[0/i] = t\).
Note that \(\refl_t\) is well-typed here because
when \(\varphi\) is true, \(p\) must be constant.
And the construction of the Swan identity type can be thought of starting with \(\refl\)
and freely adding enough compositions \emph{as well as} coercions
until the resulting type family supports \(\hcomp\) and \(\coe\).

However, Swan's construction uses the particular properties of the identity type
so that there is a minimal amount of new values added to the type.
For general higher indexed inductive types,
it is no longer obvious that a simple non-inductive constructor would suffice.
Instead, we would freely add all the coercions that can happen,
but on the index of the inductive type only.
This leads to the \(\operatorname{fcoe}\) constructor
by \citet{cavallo:2019:hit} in Cartesian cubical type theory,
and the \(\operatorname{transpX}\) constructor
by \citet{agda:2019:cubical} in Agda's de~Morgan type theory.
In this case, there would be a new normal form \(\operatorname{fcoe}\),
with some appropriate equations.

\section{Partial types and extension types}

Another common extension of cubical type theory comes from exposing more
cubical constructs that are valid in the standard cubical semantics to syntax.
This can help to internalize the types of constructs such as \(\hcomp\).

An innocuous example is the addition of interval function types
\(\mathbb{I} \to A\), which behaves essentially the same as path types,
except without the endpoint constraints.
The normal and neutral forms of this type is identical to ordinary function types.
Dependent function types \((i : \mathbb{I}) \to A(i)\) can be handled similarly.

A common pattern of reasoning in cubical type theory is
to have an element \(\varphi \vdash a : A\) that's defined only when \(\varphi\) is true,
which is intuitively specifying a subset of a cube in \(A\).
We refer to this as a \emph{partial element}.
A lot of operations in cubical type theory can be seen as taking or producing partial elements.
This leads to the idea of defining a type of partial elements:
\newcommand{\Partial}{\operatorname{Partial}}
\begin{mathpar}
\inferrule{
  \Gamma \vdash \varphi \iscof \\
  \Gamma, \varphi \vdash A \istype
}{
  \Gamma \vdash \Partial^{\varphi} (A) \istype
} \and
\inferrule{
  \Gamma, \varphi \vdash a : A
}{
  \Gamma \vdash [a] : \Partial^{\varphi} (A)
} \and
\inferrule{
  \Gamma \vdash \varphi \istrue \\
  \Gamma \vdash a : \Partial^{\varphi} (A)
}{
  \Gamma \vdash \operatorname{elim}^{\varphi}(a) : A
}
\end{mathpar}
For example, under the context \(i : \mathbb{I}\), we have a partial type
\(\Partial^{(i = 0) \lor (i = 1)} (\bool)\).
Using a case-split on the cofibration, we can define the element
\[i : \mathbb{I} \vdash \left[\begin{matrix*}
  (i=0) \hookrightarrow \btrue\\
  (i=1) \hookrightarrow \bfalse
\end{matrix*}\right] : \Partial^{(i = 0) \lor (i = 1)} (\bool).\]
We can view partial types as function types \((\varphi \istrue) \to A\)
assuming the validity of some cofibration.
From this perspective, it is also straightforward to write down
the normal and neutral forms.

However, partial types are problematic because they do not support \(\coe\) operations.
To see this, consider the partial type \(\Partial^{i=0}(\mathbf{0})\).
If \(i = 1\), then the this type has a vacuous element,
while if \(i = 0\) then the type is empty.
This means \(\coe^{1 \rightsquigarrow 0} (i \bind \Partial^{i=0}(\mathbf{0}), t)\)
would cause a contradiction.
For this reason, many cubical proof assistants have another judgment
\(\Gamma \vdash A \ispretype\) indicating \(A\)
can occur on the right of the copula, but does not support \(\hcomp\) or \(\coe\).%
\footnote{In principle we can make finer distinctions, such as
having a judgment for being able to support \(\hcomp\) and \(\coe\) separately.
This approach is taken by the RedPRL proof assistant.}

\begin{remark}
  The interval \(\mathbb{I}\) can also be considered a pretype.
  Although some proof assistants allow the formation of function (pre)types
  between pretypes, in which case \(\mathbb{I}\) cannot even be a pretype,
  because it would easily make the theory of cofibration undecidable.

  On the other hand, we can introduce the universe \(\mathbb{F}\) of cofibrations,
  similar to how \(\univ\) is a universe of types.
  \(\mathbb{F}\) is Kan, meaning it supports \(\coe\) and \(\hcomp\).
\end{remark}

We can further develop the idea by introducing a (pre)type of elements
\(a : A\) that restricts to a partial element \(\hat{a} : \Partial^{\varphi}(A)\).
This is known as the \emph{restriction type}, or cubical subtype.
\begin{mathpar}
\inferrule{
  \Gamma \vdash A \istype \\
  \Gamma, \varphi \vdash \hat{a} : A
}{
  \Gamma \vdash \{A \mid \varphi \hookrightarrow \hat{a}\} \ispretype
} \and
\inferrule{
  \Gamma \vdash a : A \\
  \Gamma, \varphi \vdash a = \hat{a} : A
}{
  \Gamma \vdash \operatorname{in}(a) : \{A \mid \varphi \hookrightarrow \hat{a}\}
} \and
\inferrule{
  \Gamma \vdash a : \{A \mid \varphi \hookrightarrow \hat{a}\}
}{
  \Gamma \vdash \operatorname{out}(a) : A
} \and
\inferrule{
  \Gamma \vdash a : \{A \mid \varphi \hookrightarrow \hat{a}\} \\
  \Gamma \vdash \varphi \istrue
}{
  \Gamma \vdash \operatorname{out}(a) = \hat{a} : A
}
\end{mathpar}
The special judgmental equality \(\operatorname{out}(a) = \hat{a}\) is very
similar to the behavior of path types.
Indeed, for the neutral form, \(\operatorname{out}\) expands the frontier of instability.
\begin{mathpar}
\inferrule{
  \Gamma \vdash A \istype \isnf \\
  \Gamma, [\varphi] \vdash \hat{a} : A \isnf
}{
  \Gamma \vdash \{A \mid \varphi \hookrightarrow \hat{a}\} \ispretype\isnf
} \and
\inferrule{
  \Gamma \vdash a : A \isnf \\
  \Gamma, \varphi \vdash a = \hat{a} : A
}{
  \Gamma \vdash \operatorname{in}(a) : \{A \mid \varphi \hookrightarrow \hat{a}\} \isnf
} \and
\inferrule{
  \Gamma \vdash a : \{A \mid \varphi \hookrightarrow \hat{a}\} \isne^{\inst{\psi}}
}{
  \Gamma \vdash \operatorname{out}(a) : A \isne^{\inst{\psi \land \varphi}}
}
\end{mathpar}

We can use these types to write \(\hcomp\) as a function
\begin{align*}
&\mathit{hcomp} :
(A : \univ)
(r~s : \mathbb{I})
(\varphi : \mathbb{F}) \\
&\quad \to (t : (i : \mathbb{I}) \to \smash{\Partial^{(i = r) \lor \varphi}} (A)) \\
&\quad \to \{A \mid (r = s) \lor \varphi \hookrightarrow \operatorname{elim}^{\varphi} t(s)\}
\end{align*}
This formulation makes some higher-order manipulations less repetitive.

The restriction type cannot support \(\hcomp\) or \(\coe\).
For example,
\[\left\{\bool \ \middle|\
\begin{aligned}
&(i = 0) \hookrightarrow \btrue \\
&(i = 1) \hookrightarrow \bfalse
\end{aligned}
\right\}\]
has the element \(\operatorname{in}(\btrue)\) when \(i = 0\),
but we cannot turn it into an element at \(i = 1\) using \(\hcomp\),
since this would imply \(\btrue = \bfalse\).
On the other hand, we can encode path types using an interval function type
together with a restriction type:
\[
\Path_A (t, t') \approx (i : \mathbb{I}) \to \left\{A \ \middle|\
\begin{aligned}
&(i = 0) \hookrightarrow t \\
&(i = 1) \hookrightarrow t'
\end{aligned}
\right\}
\]
More generally, if we combine an interval function type with
a restriction type, so that all the interval variables in the cofibration
are bound variables of the interval function type,
then the resulting combination can be given \(\hcomp\) and \(\coe\) structures.
This combination is known as \emph{extension types}.
In fact, cooltt does not contain primitive path types at all,
and chooses to encode them using extension types.

\bibliography{ref.bib}

\end{document}